\documentclass[letter]{jpsj2} 
\usepackage{epstopdf}
\title{Resistivity and upper critical field in KFe$_2$As$_2$ single crystals}

\author{Taichi \textsc{Terashima}$^{1, 3}$, Motoi \textsc{Kimata}$^{1}$, Hidetaka \textsc{Satsukawa}$^{1}$, Atsushi \textsc{Harada}$^{1}$, Kaori \textsc{Hazama}$^{1}$, Shinya \textsc{Uji}$^{1, 3}$, Hisatomo \textsc{Harima}$^{2, 3}$, Gen-Fu \textsc{Chen}$^4$, Jian-Lin \textsc{Luo}$^4$, and Nan-Lin \textsc{Wang}$^4$}

\inst{$^{1}$National Institute for Materials Science, Tsukuba, Ibaraki 305-0003\\
$^{2}$Department of Physics, Graduate School of Science, Kobe University, Kobe, Hyogo 657-8501\\
$^{3}$JST, Transformative Research-Project on Iron Pnictides (TRIP), Chiyoda, Tokyo 102-0075\\
$^{4}$Beijing National Laboratory for Condensed Matter Physics, Institute of Physics, Chinese Academy of Sciences, Beijing 100190, China}

\abst{We report resistivity measurements performed on KFe$_2$As$_2$ single crystals down to $T$ = 0.3 K and in magnetic fields up to 17.5 T.   The in-plane resistivity vs. $T$ curve has a convex shape down to $\sim$50 K and shows a $T^2$ dependence below $\sim$45 K.  The ratio of the $c$-axis to in-plane resistivities is $\sim$10 at room temperature and $\sim$40 at 4.2 K.  The superconducting upper critical field $B_{c2}(T)$ has been determined from the resistivity data: $B_{c2}^{ab}$(0) = 4.47 T and $B_{c2}^{c}$(0) = 1.25 T.  The anisotropy parameter $\Gamma$ = $B_{c2}^{ab}$/$B_{c2}^{c}$ increases with $T$ and is 6.8 at $T$ = $T_c$.  The strong curvature of  the $B_{c2}^{ab}(T)$ curve indicates the existence of strong spin paramagnetic effects.}

\kword{iron pnictide superconductor, resistivity, upper critical field, anisotropy}

\begin{document}
\maketitle


Since the discovery of superconductivity at the transition temperature $T_c = 26$ K in LaFeAsO$_{1-x}$F$_x$ by Kamihara \textit{et al}.\cite{Kamihara08JACS}, immense efforts have been put to explore possibilities of high-temperature superconductivity in iron-pnictides and related compounds.  $T_c$ has quickly been raised to 43 K by the application of pressure\cite{Takahashi08Nature} and to 54--56 K by the use of high-pressure synthesis and/or rare-earth/actinoid substitution.\cite{Kito08JPSJ, Ren08CPL, Yang08SST, Wang08EPL}  Different crystal structures have also been found to support this new class of the high-temperature superconductivity: (Ba$_{1-x}$K$_x$)Fe$_2$As$_2$ ($T_c = 38$ K), \cite{Rotter08PRL} $\alpha$-FeSe ($T_c = 8$ K),\cite{Hsu08PNAS} and LiFeAs ($T_c = 18$ K),\cite{Wang08SSC} for example, although $\alpha$-FeSe is not a pnictide.  The key ingredient of the high-temperature superconductivity seems the two-dimensional square lattice of Fe$^{2+}$ ions and it is suggested that the paring mechanism may be of magnetic origin.

(Ba$_{1-x}$K$_x$)Fe$_2$As$_2$ is a convenient system where doping dependence of superconducting and other physical properties can be studied all the way from $x=0$ to 1.  The end member BaFe$_2$As$_2$ exhibits a structural and magnetic phase transition at $T_N= 140$ K from a paramagnetic tetragonal phase to an antiferromagnetic orthorhombic phase.\cite{Rotter08PRB}  As carriers are doped, $T_N$ is suppressed and superconductivity emerges.\cite{Rotter08PRL, Rotter08ACIE}  $T_c=38$ K is achieved at $x\sim0.4$.\cite{Rotter08ACIE, Chen09EPL}   The other end member KFe$_2$As$_2$ is still superconducting but with a much lower $T_c$ of 3.8 K for polycrystals.\cite{Sasmal08PRL, Rotter08ACIE, Chen09EPL}  The continuous evolutions of  $T_c$ and the lattice parameters from $x\sim0.4$ to $x=1$ (KFe$_2$As$_2$) renders comparative studies of optimally doped $x\sim0.4$ compounds and KFe$_2$As$_2$ interesting and important.  We here report resistivity measurements performed on single crystals of KFe$_2$As$_2$ at temperatures $T$ down to 0.3 K and in magnetic fields $B$ up to 17.5 T.  We use resistivity data to determine upper critical fields $B_{c2}$.

\begin{figure}[tb]
\begin{center}
\includegraphics[width=6.5cm]{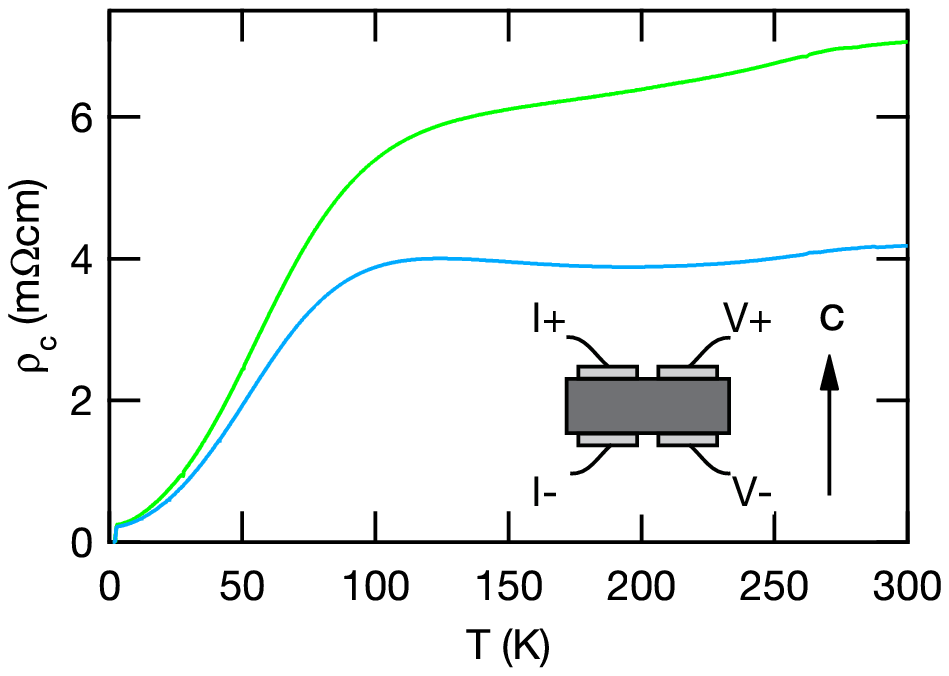}
\end{center}
\caption{(Color online) Temperature dependence of the $c$-axis resistivity of KFe$_2$As$_2$ for two samples.  The arrangement of contacts is schematically shown in the inset.}
\label{rho_c}
\end{figure}

Single crystals were grown by a flux method as described in ref.~\citen{Chen08PRB}.  The resistivity was measured by a four-contact method using low-frequency ac current ($I=0.1\sim0.3$mA, $f\sim$10 Hz).  Four gold wires were attached to (a) freshly-cleaved (001) surface(s) with conducting silver paste (for $c$-axis measurements, see the inset of Fig.~\ref{rho_c}).  A $^3$He refrigerator or a helium flow cryostat and a superconducting magnet were used to produce an experimental $T$-$B$ environment.  In in-plane resistivity measurements, the magnetic field dependence was also measured with $B$ applied perpendicular to $I $.  The field angle $\theta$ was measured from the $c$ axis.

\begin{figure}[tb]
\begin{center}
\includegraphics[width=6.5cm]{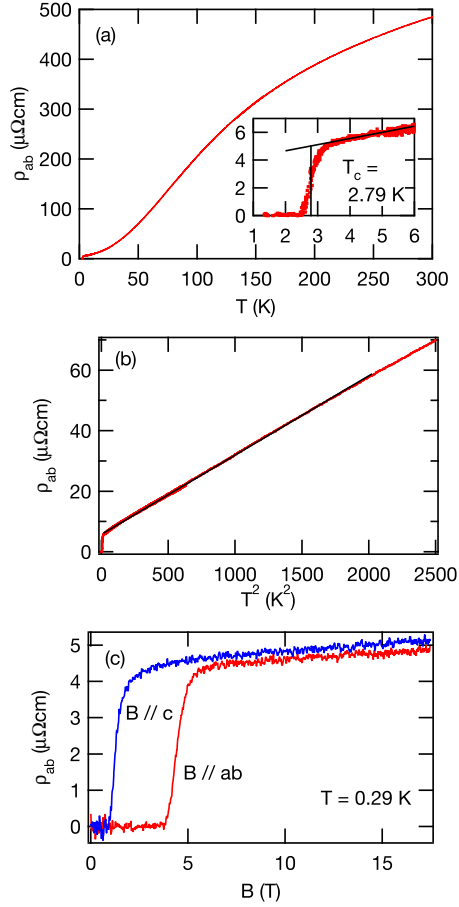}
\end{center}
\caption{(Color online) (a) Temperature dependence of the in-plane resistivity $\rho_{ab}$ of KFe$_2$As$_2$.  The inset shows a superconducting transition.  (b) Low-temperature in-plane resistivity as a function of $T^2$.  The solid line, which is almost indistinguishable from the experimental data, is a linear fit between $T$ = 4 and 45 K.  (c) Magnetic field dependence of the in-plane resistivity at $T=0.29$ K for $B \parallel c$ and $B \parallel ab$.}
\label{rho_ab}
\end{figure}

Figure~\ref{rho_ab}(a) shows the temperature dependence of the in-plane resistivity $\rho_{ab}$ at zero field.  Regarding $\rho_{ab}$(4 K) = 5.6 $\mu\Omega$cm as the residual resistivity, the residual resistivity ratio (RRR) is $\rho_{ab}$(RT)/$\rho_{ab}$(4 K) = 87.  As shown in the inset, $T_c=2.79$ K from the midpoint of the resistive transition.  This $T_c$ is lower than the polycrystal value.  The $\rho_{ab}-T$ curve shows a negative curvature, i.e., d$^2\rho_{ab}$/d$T^2 < 0$, in a wide temperature range from 300 K down to about 50 K, which is consistent with previous reports.\cite{Rotter08PRL, Sasmal08PRL, Chen09EPL}  We note that $\rho_{ab}$ exhibits a $T^2$ dependence below $\sim45$ K [Fig.~\ref{rho_ab}(b)].   The coefficient of the $T^2$ term is estimated to be $A=0.026$~$\mu\Omega$cm/K$^2$ from a fit between $T$ = 4 and 45 K.  If the Kadowaki-Woods relation can be applicable,\cite{Kadowaki86SSC} the estimated value of $A$ corresponds to the Sommerfeld coefficient $\gamma$ of 51~mJ/K$^2$mol-Fe.  This estimate appears to be too large, but there is no direct determination of $\gamma$ to compare with at the moment.  There is a phonon mechanism to explain the $T^2$ dependence of the resistivity.\cite{Kukkonen78PRB}  It is however applicable only to semimetals with small Fermi surfaces.  KFe$_2$As$_2$ is an uncompensated metal with large Fermi surface.  Therefore, the origin of the $T^2$ dependence remains unclear.

Figure~\ref{rho_ab}(c) shows $\rho_{ab}-B$ curves at $T=0.29$~K for $B \parallel c$ and $B \parallel ab$.  After the superconductivity is killed by the magnetic field, $\rho_{ab}$ increases approximately linearly with $B$ for both field directions.  Despite the small residual resistivity [$\rho_{ab}$(4 K) = 5.6 $\mu\Omega$cm] and the large RRR (= 87), which indicate the high quality of the sample, the magnetoresistivity is small.  It might be useful to recall here a textbook fact that if there is only single type of carriers, namely all carriers have the same mobility, there is no magnetoresistance.\cite{Ziman72Book}  The small magnetoresistivity might therefore indicate that KFe$_2$As$_2$ can be approximated by a single-carrier model.

Figure~\ref{rho_c} shows the temperature dependence of the $c$-axis resistivity $\rho_c$ for two samples.  The temperature dependence becomes weak above about 100 K and shows a shallow hollow around 200 K.  Based on $\rho_{ab}$ measurements on three samples and $\rho_c$ measurements on four samples, the resistivity anisotropy $\rho_c/\rho_{ab}$ is estimated to be 10 at room temperature and 40 at 4.2 K with the estimated errors of $\pm50$\% and $\pm30$\%, respectively.

\begin{figure}[tb]
\begin{center}
\includegraphics[width=6.5cm]{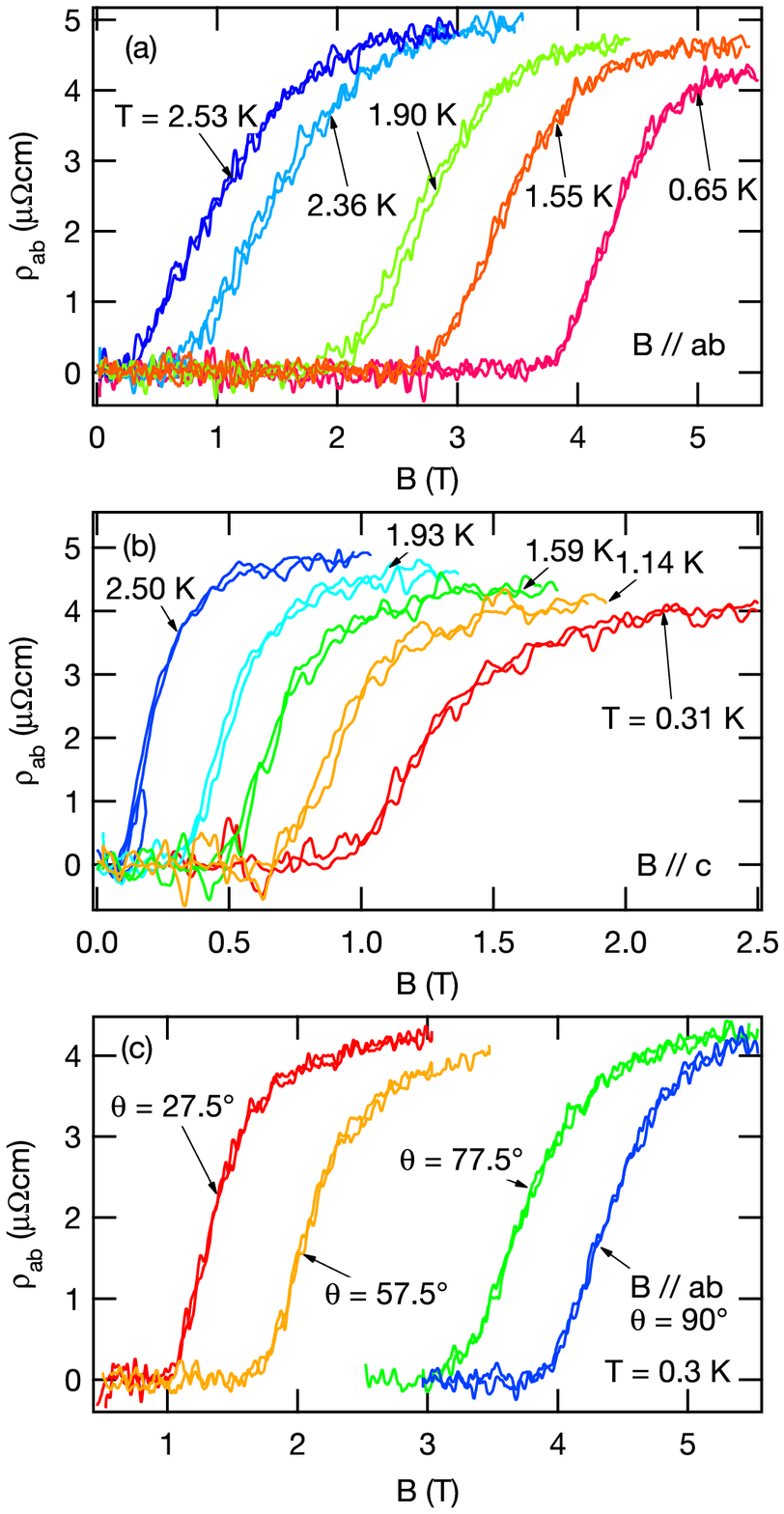}
\end{center}
\caption{(Color online) Superconducting resistive transitions at various temperatures for (a) $B \parallel ab$ and (b) $B \parallel c$ and (c) for various field directions at $T=0.3$~K.  Both increasing and decreasing field data are shown.  Because of a time constant of a lock-in amplifier, an increasing-field and the corresponding decreasing-field curve do not exactly coincide but are slightly horizontally shifted with respect to each other.  The upper critical field was estimated from the average of the midpoints of two transition curves.}
\label{transitions}
\end{figure}

\begin{figure}[tb]
\begin{center}
\includegraphics[width=6.5cm]{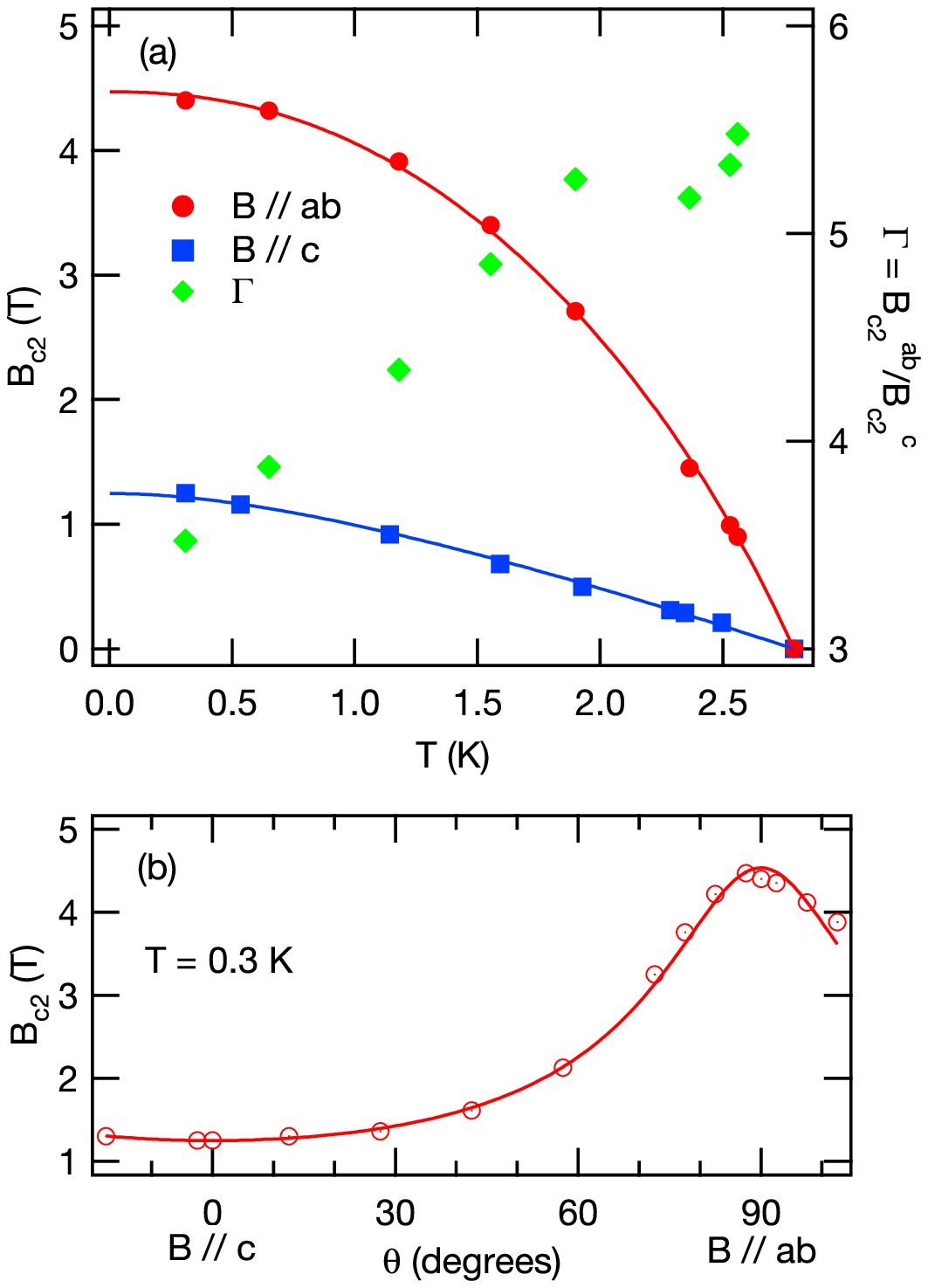}
\end{center}
\caption{(Color online) (a) Upper critical field $B_{c2}$ for $B \parallel ab$ and $B \parallel c$, and anisotropy parameter $\Gamma = B_{c2}^{ab}/B_{c2}^{c}$.   The solid curves are fits by the WHH theory (see text).  (b) Angular dependence of $B_{c2}$ at $T=0.3$~K.  The solid curve is a fit by a GL formula (see text).}
\label{Bc2}
\end{figure}

Figure~\ref{transitions} shows selected resistive ($\rho_{ab}$) transition curves for various temperatures and field directions.  From these and other data, we have determined the upper critical field $B_{c2}$ as a function of temperature for $B \parallel ab$ and $B \parallel c$ and as a function of field direction at $T=0.3$ K as shown in Fig.~\ref{Bc2}.

At the lowest measured temperature $T=0.3$ K, $B_{c2}^{ab}=4.40$ T and $B_{c2}^{c}=1.25$~T for $B \parallel ab$ and $B \parallel c$, respectively.  The anisotropy parameter $\Gamma = B_{c2}^{ab}/B_{c2}^{c}$ is 3.52.  The angular dependence of $B_{c2}$ can be fitted by a Ginzburg-Landau (GL) formula $B_{c2}(\theta) = B_{c2}^c/(\cos^2\theta + \Gamma^{-2}\sin^2\theta)$ with $\Gamma=3.63$ as shown by the solid curve in Fig.~\ref{Bc2}(b).\cite{Decroux82Book}  This fit is, however, only phenomenological since the formula is derived for the orbital critical field $B_{c2}^*$ and $B_{c2}$ of KFe$_2$As$_2$ is largely affected by spin paramagnetic effects as shown below.  The anisotropy parameter $\Gamma$ increases to $\sim$5.5 as $T$ approaches $T_c$ as shown in Fig.~\ref{Bc2}(a).

The initial slope d$B_{c2}$/d$T$ at $T = T_c$ is estimated to be $-3.8$ and $-0.71$ T/K for $B \parallel ab$ and $B \parallel c$, respectively, from the data points for $T > 2.4$ K.  These values can be converted to the orbital critical fields $B_{c2}^*(0)$ of 7.4 and 1.4 T, respectively, using a Werthamer-Helfand-Hohenberg (WHH) formula.\cite{Werthamer66PRB}   The Maki parameter $\alpha$ can also be estimated to be 2.0 and 0.38 for $B \parallel ab$ and $B \parallel c$, respectively.  The large Maki parameter and the fact that the observed $B_{c2}$ at $T=0.3$ K for $B \parallel ab$ is much smaller than the orbital critical field indicate that there are strong spin paramagnetic effects for $B \parallel ab$.  The above-mentioned temperature dependence of $\Gamma$ can be explained by the suppression of $B_{c2}$ for $B \parallel ab$ at low temperatures by the spin paramagnetic effects.  We note that the $B_{c2}$ curve of nearly optimally doped (Ba,K)Fe$_2$As$_2$ also shows clear convex curvature, a sign of the strong spin paramagnetic effects, only for $B \parallel ab$.\cite{Yuan09Nature, Altarawneh08PRB}

In order to determine the anisotropy parameter $\Gamma$ at $T=T_c$, which is directly connected with the effective mass anisotropy, the $B_{c2}$ data have been fitted by the WHH theory that incorporates both orbital and spin paramagnetic effects.\cite{Werthamer66PRB}  The parameters of the theory are the Maki parameter $\alpha$ and the spin-orbit parameter $\lambda_{so}$.  In the present fitting, $\lambda_{so}$ was fixed to be $\infty$ for $B \parallel c$, namely the spin paramagnetic effects were neglected for $B \parallel c$, and $T_c=2.79$~K was fixed for both field directions.  The resultant fits are excellent as shown by the solid lines in Fig.~\ref{Bc2}(a) and the fitted parameters are listed in Table~\ref{t1}.  From the determined $\alpha$ values, $\Gamma=(m_c/m_{ab})^{(1/2)}=6.8$ at $T=T_c$, where $m_c$ and $m_{ab}$ are the effective masses of the carriers along the $c$ axis and in the $ab$ plane, respectively.  Table~\ref{t1} also shows $B_{c2}(0)$, $B_{c2}^*(0)$, and the coherence length $\xi$ derived from the fitted parameters, and the paramagnetic critical field $B_{po}=1.84T_c$ [Tesla]. 

One might ask why $\Gamma$ in KFe$_2$As$_2$ is so large.  The determined $\Gamma$ of 6.8 is considerably larger than a typical value of $\Gamma \sim 2$ at $T=T_c$ found in nearly optimally doped (Ba, K)Fe$_2$As$_2$.\cite{Yuan09Nature, Ni08PRB, Wang08PRB}  It is however not inconsistent with the fairly large resistivity anisotropy of $\sim40$.  Electronic band structure calculations for tetragonal paramagnetic BaFe$_2$As$_2$ with the experimental lattice parameters predict quasi-two-dimensional Fermi surface consisting of electron and hole cylinders.\cite{Nekrasov08JETPLett, Harima08PC}  Within the rigid-band approximation, the Fermi surface of KFe$_2$As$_2$ is expected to be quasi-two-dimensional consisting of hole cylinders centered at the $\Gamma$ point in the Brillouin zone.  In this view, the observed large $B_{c2}$ and resistivity anisotropies of KFe$_2$As$_2$ are reasonable.  This reveals that the question to be asked is, actually, why the anisotropy of optimally doped (Ba, K)Fe$_2$As$_2$ is small.

\begin{table}[tb]
\caption{Superconducting parameters from the WHH fits.}
\label{t1}
\begin{tabular}{ccc}
\hline
 & $B \parallel ab$ & $B \parallel c$ \\
\hline
$\alpha$ & 2.30 & 0.340 \\
$\lambda_{so}$ & 0.36 & $\infty$ \\
$T_c$ & \multicolumn{2}{c}{2.79 K} \\
\\
$B_{c2}(0)$ & 4.47 T & 1.25 T \\
$B_{c2}^*(0)$ & 8.44 T & 1.25 T \\
$B_{po}$ & \multicolumn{2}{c}{5.13 T} \\
\\
\multicolumn{3}{c}{$\xi_{ab}=16.3$~nm, $\xi_{c}=2.45$~nm} \\
\hline
\end{tabular}
\end{table}

In summary, we have measured resistivity of KFe$_2$As$_2$ single crystals and have determined the upper critical field.  The $\rho_{ab}(T)$ curve exhibits a convex curvature down to $\sim$50 K and a $T^2$ dependence below $\sim$45 K with a fairly large coefficient $A$ = 0.026 $\mu\Omega$cm/K$^2$.  The resistivity anisotropy $\rho_c/\rho_{ab}$ is $\sim$10 at room temperature and $\sim$40 at 4.2 K.  The $B_{c2}(T)$ curves are fitted by the WHH theory very well.  $B_{c2}^{ab}$(0) and $B_{c2}^{c}$(0) are estimated to be 4.47 T and 1.25 T, respectively.  The anisotropy parameter $\Gamma$ increases from 3.5 at $T$ = 0 to 6.8 at $T$ = $T_c$.  The value at $T_c$ is much larger than that found in nearly optimally doped (Ba, K)Fe$_2$As$_2$, but is consistent with the expected quasi two-dimensional electronic structure, along with the large resistivity anisotropy.  The strong spin paramagnetic effects are found for $B\parallel ab$, which is similar to the observations in nearly optimally doped (Ba, K)Fe$_2$As$_2$.



\end{document}